\documentclass[fp,twocolumn,shortnote]{jpsj3}
\usepackage{txfonts}
\usepackage{color}

\title{Thermally Activated Motion of Sodium Cations in Insulating Parent Low-Silica X Zeolite
}

\author{
Mutsuo Igarashi$^1$\thanks{igarashi@elc.gunma-ct.ac.jp}, 
Peter Jegli\v{c}$^{2}$\thanks{peter.jeglic@ijs.si},
Tadej Me\v{z}nar\v{s}i\v{c}$^{2}$, 
Takehito Nakano$^3$, 
Yasuo Nozue$^3$,  
Naohiro Watanabe$^1$,
Denis Ar\v{c}on$^{2,4}$
}

\inst{
$^1$Gunma College, National Institute of Technology, Toribamachi 580, Maebashi 371-8530, Gunma, Japan \\
$^2$Jo\v{z}ef Stefan Institute, Jamova 39, 1000 Ljubljana, Slovenia \\
$^3$Department of Physics, Graduate School of Science, Osaka University, Toyonaka 560-0043, Osaka, Japan \\
$^4$Faculty of Mathematics and Physics, University of Ljubljana, Jadranska 19, 1000 Ljubljana, Slovenia
}

\abst{
We report a $^{23}$Na spin-lattice relaxation rate, $T_1^{-1}$, in low-silica X zeolite.
$T_1^{-1}$ follows multiple BPP-type behavior as a result of thermal motion of sodium cations in insulating material.
The estimated lowest activation energy of 15~meV is much lower than 100~meV observed previously for sodium motion in heavily Na-loaded samples and is most likely attributed to short-distance jumps of sodium cations between sites within the same supercage.
}

\begin{document}
\maketitle
\vspace{5mm}

Na-type low-silica X (LSX) zeolite loaded with guest Na atoms represents one of the first examples of an elusive metallic behavior in the zeolite systems.\cite{Nakano_T_JPCS2010,Igarashi_M_SREP2016}
The emergence of a Drude peak in optical reflectivity and finite resistivity at low temperature were demonstrated on the macroscopic scale only at high loading levels of guest sodium atoms ($n>14.2$).\cite{Nakano_T_JPCS2010}
On the other hand, a continuous (crossover like) evolution from insulating to metallic ground state has been observed with a local-probe nuclear magnetic resonance (NMR) spectroscopy by extracting the electron density of states (DOS) from low-temperature spin-lattice relaxation data.\cite{Igarashi_M_SREP2016}
The observed crossover in alkali-doped zeolites is a result of a complex loading-level dependence of electronic correlations, disorder and electron-phonon coupling.

\par

In alkali-doped zeolites the spin-lattice relaxation rate, $T_1^{-1}$, is dominated by strong fluctuations of local magnetic fields and electric field gradients originating from large amplitude atomic motion,\cite{Heinmaa_I_CPL2000,Igarashi_M_PRB2013} which can be modeled within the so-called Bloembergen-Purcell-Pound (BPP) relaxation theory.\cite{BPP_PR1948}
In sodium doped LSX zeolites the atomic motion is frozen on the NMR time scale at very low temperatures and thus in this temperature range other relaxation mechanisms begin to dominate. 
For instance, below 25~K, $^{23}$Na $T_1^{-1}$ shows Korringa behavior for $n\geq 14.2$, thus proving the metallic ground state.\cite{Igarashi_M_SREP2016}
Surprisingly, a small portion of DOS at the Fermi level persists deep into the insulating state ($9.4 \leq n\leq 11.6$), where the optical measurements show no Drude peak. 
In order to elucidate the nature of atomic motion, this work focuses on parent ($n=0$) sample of sodium LSX zeolite.

\par

The parent or non-loaded sodium LSX zeolite has a chemical formula Na$_{12}$Al$_{12}$Si$_{12}$O$_{48}$ (abbreviated Na$_{12}$-LSX) with twelve sodium cations in the structure, which are required for charge compensation of the aluminosilicate framework.
When Na$_{12}$-LSX is exposed to Na vapour, a controlled amount of $n$ sodium atoms per zeolite supercage\cite{Nozue_Y_JPCS2012} can be loaded, yielding a composition Na$_n$/Na$_{12}$-LSX.
As shown below, the thermal motion of charge compensating sodium cations fully explains the observed $^{23}$Na spin-lattice relaxation in the non-loaded compound without any hints of residual conduction electron contribution.
Present result thus sets the benchmark for the $^{23}$Na $T_1$ against which all other relaxation mechanisms in Na$_n$/Na$_{12}$-LSX samples have to be evaluated.

\begin{figure} [b!]
\includegraphics[width=7.5cm]{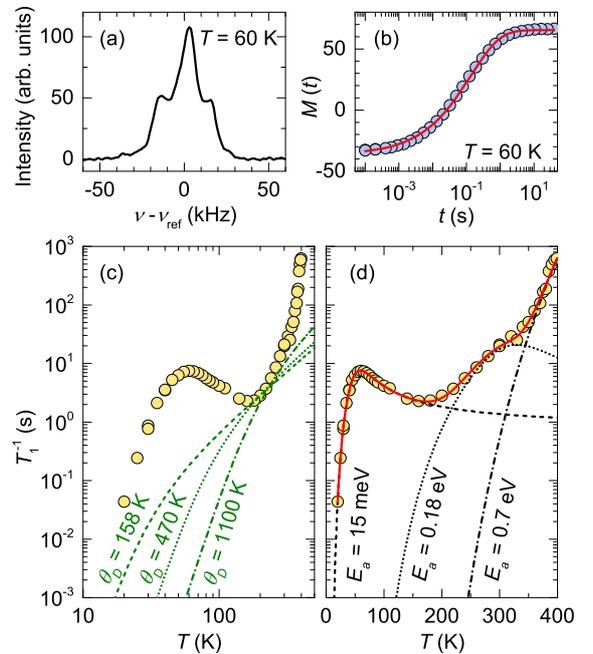}
\centering
\caption{(color online). 
(a) The $^{23}$Na NMR spectrum ($\nu_{\rm ref}=105.87$~MHz) and (b) magnetization recovery at 60 K fitted with $\alpha=0.46$ (see text for details).
(c) Hypothetical Raman contributions to experimental $^{23}$Na $T_1^{-1}$ for different values of Debye temperatures (green dashed, dotted and dash-dotted lines correspond to $\theta_D=$ 158~K, 470~K and 1100~K, respectively).
(d) Simulated temperature dependence (red solid line) of $^{23}$Na $T_1^{-1}$ (yellow circles) modeled with three BPP contributions (dashed line: $E_{a1}=15.2(5)$~meV, $\tau_{01}=7.6(6)\times 10^{-11}$~s, $C_1=1.01(5)\times 10^{10}$~s$^{-2}$, dotted line: $E_{a2}=0.18(2)$~eV, $\tau_{02}=3(2)\times 10^{-12}$~s, $C_2=3(2)\times 10^{10}$~s$^{-2}$, dash-dotted line: $E_{a3}=0.7(3)$~eV, $\tau_{03}=2(1)\times 10^{-17}$~s and $C_3=9(8)\times 10^{12}$~s$^{-2}$).

}
\label{fig1}
\end{figure}

\par

For the purpose of NMR measurements the non-loaded, just dehydrated, Na-LSX sample was sealed in a quartz ampule\cite{Nozue_Y_JPCS2012} and placed in a magnetic field of 9.4~T.
Fig.~\ref{fig1}(a) shows a representative $^{23}$Na (nuclear spin $I =3/2$)  NMR spectrum of Na-LSX measured at 60~K.
The complex lineshape is a superposition of multiple overlapping second-order quadrupolarly perturbed lineshapes. 
These are a result of crystallographically distinct sodium cation sites with zero hyperfine shift but with different quadrupolar coupling constants.\cite{Feuerstein_M_2006}
As a consequence, the $^{23}$Na $T_1$ measured with the inversion recovery technique is a mean of distribution of spin-lattice relaxation times over different sodium sites.
In addition, for each sodium site the nuclear magnetization recovery follows $M(t)\propto a_1 {\rm exp}(-b_1 t/T_1)+a_2 {\rm exp}(-b_2 t/T_1)$, with two exponential components with coefficients $a_1$, $a_2$, $b_1$ and $b_2$ depending on the initial condition and relaxation origin.\cite{Andrew_ER_1961}
To take into account the distribution of $T_1$ originating from both the multitude of cation sites and the peculiarity of $I=3/2$ nuclei, we model the magnetization recovery curves with $M(t) \propto {\rm exp} [-(t/T_1)^\alpha]$, the so-called stretched exponential form.\cite{Johnston_DC_PRB2006}
The stretched exponent $\alpha$  is a measure of the width of $T_1$ distribution.
Fig.~\ref{fig1}(b) shows an example of magnetization recovery curve measured at 60~K with strikingly small $\alpha = 0.46$.
The value of the stretched exponent implies a distribution of $T_1$ that spreads over two orders of magnitude.\cite{Johnston_DC_PRB2006} 
It is important to note that the value of $\alpha$ remains approximately constant across the whole 20-400~K temperature range. 

\par

As evidenced in Fig.~\ref{fig1}(c), $^{23}$Na $T_1^{-1}$ decreases exponentially with decreasing temperature below 60~K, ruling out the possibility of any residual electron (metallic) contribution.
However, a nuclear quadrupolar Raman (phonon-like) process might still be relevant in Na-LSX. 
We try to estimate the temperature dependence of Raman contribution to the spin-lattice relaxation of the form\cite{Abragam}
\begin{equation}
T_1^{-1} (T) \simeq R \left( \frac{T}{\theta_D} \right) ^7 \int _{0} ^{\theta_D /T} \frac{e ^{x} x^6}{(e^{x}-1)^2} dx, 
\label{QRT1}
\end{equation}
\noindent
where $R$ is a constant governed by the strength of quadrupolar interaction and $\theta_D$ is the Debye temperature. 
Eq.~(\ref{QRT1}) gives $T_1^{-1}$ proportional to $T^2$ for $T>\theta_D/2$ and to $T^7$ for $T<0.02 ~\theta_D$. 
Unfortunately, the value of $\theta_D$ is not known for Na-LSX or related zeolite materials.
The range of possible $\theta_D$ values can be estimated from two limiting cases.
In the limit of soft bonding, the bulk Na with $\theta_D = 158$~K  
gives the lower value\cite{Kittel}.
On the other hand, in the most rigid limit the upper value for $\theta_D$ may be given from the similarity between the Al-Si-O zeolite framework and SiO$_2$.
The reported values for SiO$_2$ vary in a wide range from 250~K to 1100~K.\cite{Lord_RC_1957,Striefler_ME_1975}
We plot the Raman contributions calculated from Eq. (\ref{QRT1}) for different values of $\theta_D$ in Fig.~\ref{fig1}(c) by taking into account a constraint that the maximal contribution cannot exceed the measured $T_1^{-1}$ at any temperature.
We choose the value of $R$ in such a way that the calculated Raman contribution passes through the minimum in $T_1^{-1}$ around 180~K.
Clearly, the Raman contribution is negligibly small at low temperatures in Na-LSX.

\par 

A different approach is needed to explain the temperature dependence of $^{23}$Na $T_1^{-1}$.
The local maximum in $T_1^{-1}$ at 60~K is a hallmark of BPP-type relaxation,\cite{BPP_PR1948} which is described by $T_1^{-1}= C\tau_c/(1+\omega^2\tau_c^2)$.
Here $\omega$ is the $^{23}$Na Larmor angular frequency, $\tau_c$ is the correlation time for  local field fluctuations at the nucleus due to atomic motion and $C$ is a measure of  coupling to the relaxation source.
Next we assume that the atomic motion exhibits thermally activated Arrhenius behavior: $\tau_c = \tau_0 \exp \left[ E_a/(k_B T) \right]$,\cite{Arrhenius_PRB1984} with activation energy $E_a$ for the sodium jumps between two crystallographic sites.
$\tau_0$ is a constant and $k_B$ is the Boltzmann constant.
The dashed line in Fig.~\ref{fig1}(d) represents a fit to the low-temperature experimental data.
The activation energy of 15~meV is much lower than the previously reported values of around 100~meV found at high loading levels,\cite{Igarashi_M_PRB2013,Igarashi_M_SREP2016} which suggests that a different type of Na cation jumps is involved.
{Alternatively, the low-temperature local maximum in $T_1^{-1}$ at 60~K could arise from the sodium cation rattling.
However, by adopting the model of the rattling within an anharmonic double well potential in $\beta$-pyrochlore structure,\cite{Yoshida_2007} we obtain an activation energy of about 16~meV (180~K), which is unreasonably high for the rattling.

\par

Na-LSX has two types of cages, smaller $\beta$ cages and larger supercages with an inner diameter of $\sim$7~$\AA$ and $\sim$13~$\AA$, respectively, both forming a diamond-type super-structure.\cite{Kien_LM_2015}
They are mutually connected through ``windows'' with diameters ranging from 2.8~{\AA} to 8~{\AA}.\cite{Kien_LM_2015}
The $\beta$ cage sites are expected to be fully occupied with sodium cations, leaving no space for the intra-cage sodium jumps.\cite{Nakano_T_JPCS2010}.
Therefore, the low-temperature atomic motion observed in Na-LSX most likely involves sodium cations occupying some of the available sites within the supercage.
According to our full analysis of $^{23}$Na $T_1^{-1}$ shown in Fig.~\ref{fig1}(d) two additional modes of sodium cation motion occur at higher temperatures with much higher activation energies, $E_{a2}=0.18(2)$~eV and $E_{a3}=0.7(3)$~eV.
These extracted parameters are less reliable due to lack of clear BPP maxima.
Nevertheless, the observed high-temperature motion with high activation energies can involve sodium cations, which are originally residing at the $\beta$ cage sites, but can occasionally jump through framework windows to the neighboring cages. 
We conclude that, because of negligibility of the Raman contribution, the observed $T_1^{-1}$ is fully explained solely by thermal sodium cation motion.
This motion is much more complicated, as thought previously as it comprises low activation-energy jumps within the large supercages as well high activation-energy jumps between neighboring $\beta$ cages and/or supercages.

\par

In this work, we have investigated the $^{23}$Na spin-lattice relaxation rate of non-loaded Na-LSX zeolite.
In comparison to Na$_{9.4}$/Na$_{12}$-LSX that still showed some residual DOS at low temperatures,\cite{Igarashi_M_SREP2016} the Na-LSX data reveal no remnants of metallicity. 
This confirms that the non-loaded compound is deep in the insulating ground state.
The $^{23}$Na $T_1^{-1}$ data reveal three types of sodium cation motion present in the parent Na-LSX, meaning that the atomic motion does not appear only in loaded Na$_{n}$/Na$_{12}$-LSX samples.
The low-energy motion has been attributed to cations occupying the supercage sites, whereas two high-energy motions most likely involve large-amplitude jumps of sodium cations originally occupying the $\beta$ cage sites.

\begin{acknowledgment}

Authors acknowledge fruitful discussions with Katsumi Tanigaki.
This study was partially supported by Grants-in-Aid for Scientific Research (KAKENHI) [Grants No. 24244059(A), 16K05462(C), and No. 26400334(C)] from Japan Society for the Promotion of Science.

\end{acknowledgment}

\end{document}